\documentclass[12pt]{article}

\usepackage{amsmath}
\usepackage{amssymb}
\usepackage{epsfig}
\usepackage{psfrag}
\usepackage{color}

\usepackage{geometry}
\newcommand{\comments}[1]{}

\makeatletter
\@addtoreset{equation}{section}
\makeatother

\newcommand{\cqfd}{{\unskip\kern 6pt\penalty 500
\raise -2pt\hbox{\vrule\vbox to 6pt{\hrule width 6pt
\vfill\hrule}\vrule}\par}}

\begin{document}
\title{Condensation with two constraints and disordered Discrete Non Linear Schr\"odinger breathers}
\author{J.~Barr\'e$^1$,\quad julien.barre@univ-orleans.fr\\
L.~Mangeolle$^2$,\quad leo.mangeolle@ens-lyon.fr\\ } 
\footnotetext[1]{Institut Denis Poisson, Universit\'e d'Orl\'eans, Universit\'e de Tours et CNRS, Orl\'eans, France, and Institut Universitaire de France, Paris, France}
\footnotetext[2]{ENS Lyon}
\date{\today}
\maketitle
\begin{abstract}
Motivated by the study of breathers in the disordered Discrete Non Linear Schr\" odinger equation, we study the uniform probability over the intersection of a simplex and an ellipsoid in $n$ dimensions, with quenched disorder in the definition of either the simplex or the ellipsoid. Unless the disorder is too strong, the phase diagram looks like the one without disorder, with a transition separating a fluid phase, where all variables have the same order of magnitude, and a condensed phase, where one variable is much larger than the others. We then show that the condensed phase exhibits "intermediate symmetry breaking": the site hosting the condensate is chosen neither uniformly at random, nor is it fixed by the disorder realization. In particular, the model mimicking the well-studied Discrete Non Linear Schr\"odinger model with frequency disorder shows a very weak symmetry breaking: all variables have a sizable probability to host the condensate (i.e. a breather in a DNLS setting), but its localization is still biased towards variables with a large linear frequency. Throughout the article, our heuristic arguments are complemented with direct Monte Carlo simulations.  
\end{abstract}

\section{Introduction}
We start from the following idealized problem: consider the uniform distribution on the surface defined by
\begin{eqnarray}
\begin{cases}
\forall i=1,\ldots,n~,~x_i \geq 0  \\
\sum_{i=1}^n x_i = n m_1  \\
\sum_{i=1}^n x_i^2 = n m_2 
\end{cases}
\label{eq:1}
\end{eqnarray}
This surface is the intersection of a simplex and a sphere. Take a random point on this surface; the question is: what does it look like? In particular, what is the probability distribution of its coordinates? This seemingly simple question has an interesting answer: the probability distribution of the coordinates undergoes a condensation phenomenon when $m_2/m_1^2 >2$. This was first noticed in the context of the Discrete Non Linear Schr\"odinger (DNLS) equation \cite{Rasmussen00,Johansson04,Rumpf04,Rumpf08}, where it explains how 
large localized nonlinear structures called "breathers" can be thermally excited. It was later studied in more details, including generalizations \cite{Filiasi14,Szavits14a,Szavits14b}. A rigorous proof of the condensation phenomenon in \eqref{eq:1} has also been provided \cite{Chaterjee10}, and the setting of \eqref{eq:1} has been applied to give a statistical description of dark matter halos \cite{Carron13}.
Condensation phenomena are of course much more general than \eqref{eq:1}; in particular, often in connection with the Zero Range Process, there is a huge body of works considering the case where the $x_i$ are distributed according to a heavy-tailed product density with a linear constraint on their sum: we can only refer to a few papers here, because the literature is truly enormous \cite{Grosskinsky03,Evans05,Majumdar05}; \cite{Majumdar_review} provides a review. \cite{Filiasi14,Szavits14a,Szavits14b} explore in detail situations with two constraints, hence closer to \eqref{eq:1}.

The DNLS equation reads, for complex dynamical variables $\phi_k(t)$:
\begin{equation}
i\partial_t \phi_k = \beta_k |\phi_k|^2 \phi_k +\omega_k \phi_k -\kappa(\phi_{k-1}+\phi_{k+1}),
\label{eq:ddnls}
\end{equation}
where $\omega_k$ is the onsite frequency, and $\beta_k$ the onsite nonlinearity; $\kappa$ is the coupling between neighboring sites. The homogeneous case corresponds to $\beta_k,\omega_k$ independent of $k$, and the case where either the $\beta_k$ or the $\omega_k$, or both, are quenched random variables will be refered to as disordered DNLS. \eqref{eq:ddnls} has two conserved quantities, the norm $I$ and the Hamiltonian $H$:
\begin{eqnarray}
I&=& \sum_k |\phi_k|^2 \label{eq:norm}\\
H&=& \sum_k\left(\omega_k |\phi_k|^2 +\frac{\beta_k}{2}|\phi_k|^4 -\kappa(\phi_k\phi_{k+1}^\ast+\phi_k^\ast\phi_{k+1})\right). \label{eq:H}
\end{eqnarray}
Problem \eqref{eq:1} stems from the equilibrium microcanonical analysis of \eqref{eq:ddnls} in the homogeneous case, taking into account the two conserved quantities \eqref{eq:norm} and \eqref{eq:H}, and neglecting the coupling term \footnote{A straightforward change of variables is also needed here.}.
The homogeneous DNLS equation and its variants are used to model a wide variety of phenomena (see for instance \cite{Eilbeck02} for a review). In many applications however, the disordered version \eqref{eq:ddnls} shows up, and a large literature is devoted to it, including many studies of discrete breathers in a disordered context (for instance \cite{Feddersen91,Molina94,Molina98,Rasmussen99,Kopidakis99,Kopidakis00a,Kopidakis00b,Gupta01,Kottos11,Flach14}, again we cannot be exhaustive here; \cite{Flach08} provides a review on discrete breathers). However, the main emphasis in this literature seems to be on elucidating the interplay between non linearity, which lies at the heart of breather formation, and Anderson localization; as a consequence, to our knowledge the condensation transition for disordered DNLS has not been studied with an equilibrium statistical mechanics point of view. 
Since the simplified model \eqref{eq:1} has proved very useful to qualitatively understand the statistical mechanics of the homogeneous DNLS equation, 
our goal is to study several disordered versions of \eqref{eq:1}, with possible applications to disordered DNLS models in mind:
\begin{eqnarray}
\begin{array}{ll}
{\rm Model~I} & \begin{cases}\forall i=1,\ldots,n~,~x_i \geq 0 \\ \sum_{i=1}^n \alpha_i x_i = n m_1\\ \sum_{i=1}^n x_i^2 = n m_2  \end{cases}
\end{array}
\label{eq:model1}
\end{eqnarray}
\begin{eqnarray}
\begin{array}{ll}
{\rm Model~II} & \begin{cases}\forall i=1,\ldots,n~,~x_i \geq 0 \\ \sum_{i=1}^n x_i = n m_1\\ \sum_{i=1}^n \beta_i x_i^2 = n m_2  \end{cases}
\end{array}
\label{eq:model2}
\end{eqnarray}
\begin{eqnarray}
\begin{array}{ll}
{\rm Model~III} & \begin{cases}\forall i=1,\ldots,n~,~x_i \geq 0 \\ \sum_{i=1}^n  x_i = n m_1 \\ \sum_{i=1}^n \omega_i x_i +\sum_{i=1}^n x_i^2 = n m_2  \end{cases}
\end{array}
\label{eq:model3}
\end{eqnarray}
where the $\alpha_i,\beta_i$ and $\omega_i$ are quenched random variables with a known distribution. 
Model II can be related to a DNLS system with random on site nonlinearities \cite{Molina94,Molina98}; model III can be related to the widely studied case of random 
on site frequencies (see \cite{Rasmussen99} for instance). While model I cannot be directly related to a disordered DNLS equation, it is also a natural generalization of \eqref{eq:1}: it represents the intersection of a random direction hyperplane with a sphere (in the positive quadrant). 

We shall ask two types of questions on these models: first, how is the phase diagram modified by the disorder? In the condensed phase, one coordinate $x_i$ is much larger than the others; without disorder, the site hosting this condensate is obviously chosen uniformly at random among  all sites. Hence the second question: does the disorder induce a selection of the site hosting the condensate?  

There is a large literature dealing with condensation phenomena in models with some heterogeneity, or randomness (see for instance \cite{Evans96,Krug96,Jain03,Ferrari07,Grosskinsky08,Grosskinsky11,Godreche12,Chleboun14,Corberi15}). The stationary measure of these models is often a product measure with one single constraint, representing particles conservation. \cite{Godreche12} in particular considers in this "one-constraint" setting a type of randomness similar to ours, and finds an instance of "intermediate symmetry breaking", interpolating between spontaneous symmetry breaking, where the site hosting the condensate is chosen uniformly at random (this is the case in the absence of disorder), and explicit symmetry breaking, where the hosting site is 
deterministic once the disorder is fixed. The occurrence of this scenario has been rigorously proved in \cite{Mailler16}. We will see that this phenomenology is also present in the two-constraints setting of models I, II and III.

The article is organized as follows: in section \ref{sec:phase_diagram} we investigate how the transition between fluid (without breather) and condensed (with a  breather) phases is modified by the disorder. Our main result here is that while a weak disorder does not bring qualitative changes, the transition may disappear in presence of a strong enough disorder; here, "strong" means that some moment of the quenched random variable $\alpha,\beta$ or $\omega$ diverges. In section \ref{sec:condensed}, we investigate the selection of the hosting site, and find in general an intermediate symmetry breaking scenario. In particular, in the case of \eqref{eq:model3}, which mimicks a DNLS equation with random on site frequencies, the symmetry breaking is very weak: all sites have a sizable probability of hosting the condensate, but this probability is not uniform: it is biased towards high onsite frequency sites.

\section{The phase diagram}
\label{sec:phase_diagram}
We would like to compute the volume of the hypersurfaces defined by \eqref{eq:model1}, \eqref{eq:model2}, \eqref{eq:model3}: we shall call this the microcanonical problem. 
We will first study it in the grand canonical ensemble, which will provide the solution to the microcanonical problem whenever the ensembles are equivalent. 

\subsection{Model I}
We need some hypotheses on the random variables $\alpha_i$: we assume they are independent and identically distributed, positive, with finite expectation. Without loss of generality, we may assume that this expectation is $1$; this will facilitate the comparison with the homogeneous case where $\alpha_i=1,~\forall i$. 
The grand canonical partition function reads
\begin{eqnarray}
Z_n(\lambda,\beta) &=& \int \Pi dx_i e^{-\lambda \sum \alpha_i x_i -\mu \sum x_i^2} \nonumber \\
&=& \Pi_{i=1}^n z(\lambda \alpha_i,\mu) \nonumber 
\end{eqnarray}
where 
\[
z(\lambda,\mu) = \int_0^\infty e^{-\lambda x-\mu x^2}dx 
\]
To obtain the microcanonical distribution, the parameters $\lambda$ and $\mu$ have to be determined as solutions of the equations
\begin{eqnarray}
-\frac1n \partial_\lambda \ln Z_n(\lambda,\mu) &=& m_1 \\
-\frac1n \partial_\mu \ln Z_n(\lambda,\mu) &=& m_2 
\end{eqnarray}
This yields
\begin{eqnarray}
\frac1n \sum_{i=1}^n \alpha_i\frac{\int_0^\infty xe^{-\lambda\alpha_i x-\mu x^2}dx}{z(\lambda\alpha_i,\mu)} &=& m_1 
\label{eq:syst_a_resoudre0a} \\
\frac1n \sum_{i=1}^n \frac{\int_0^\infty x^2e^{-\lambda\alpha_i x-\mu x^2}dx}{z(\lambda\alpha_i,\mu)} &=& m_2
\label{eq:syst_a_resoudre0b}
\end{eqnarray}
Introducing $a=m_2/m_1^2$, $\tilde{\lambda}=\lambda m_1$ and $\tilde{\mu}=\mu m_1^2$, this can be rewritten (dropping the $\tilde{}$ for convenience)
\begin{eqnarray}
\varphi_1(\lambda,\mu,\{\alpha_i\})= \frac1n \sum_{i=1}^n \alpha_i\frac{\int_0^\infty xe^{-\lambda\alpha_i x-\mu x^2}dx}{z(\lambda\alpha_i,\mu)} &=& 1 
\label{eq:syst_a_resoudre1}\\
\varphi_2(\lambda,\mu,\{\alpha_i\})=\frac1n \sum_{i=1}^n \frac{\int_0^\infty x^2e^{-\lambda\alpha_i x-\mu x^2}dx}{z(\lambda\alpha_i,\mu)} &=& a
\label{eq:syst_a_resoudre2}
\end{eqnarray}
We thus look for $(\lambda,\mu)$ solution to \eqref{eq:syst_a_resoudre1}-\eqref{eq:syst_a_resoudre2}, with $\mu >0$ or $\mu=0,\lambda>0$.  
In appendix 1, we show that no such solution exists when $a$ is large enough. More precisely, we prove that under the constraint $\varphi_1(\lambda,\mu)=1$, the function $\varphi_2$ reaches its maximum for $(\lambda=1,\mu=0)$, and this maximum is
\begin{equation}
a_c^{(I)} = 2\frac1n \sum_{i=1}^n \frac{1}{\alpha_i^2} \underset{n\to \infty} {\to}2\mathbb{E}_\alpha\left[ \frac{1}{\alpha^2}\right], 
\label{eq:acI}
\end{equation}
where $\mathbb{E}_\alpha$ denotes the expectation with respect to the quenched disorder. \eqref{eq:acI} provides the transition line between the "fluid" and the "condensed" phases.
For $a<a_c^{(I)}$, \eqref{eq:syst_a_resoudre0a}-\eqref{eq:syst_a_resoudre0b} has a unique solution, and grand canonical and microcanonical ensembles are equivalent: this is usually called the "fluid phase". Denoting $(\lambda^\ast(m_1,m_2),\mu^\ast(m_1,m_2))$ the solution of \eqref{eq:syst_a_resoudre0a}-\eqref{eq:syst_a_resoudre0b}, the probability distribution of site $i$ is given, in the large $n$ limit, by
\begin{equation}
p_i(x) = \frac{e^{-\lambda^\ast(m_1,m_2) \alpha_i x-\mu^\ast(m_1,m_2) x^2}}{z(\lambda\alpha_i,\mu)},
\label{eq:gdcano}
\end{equation}
and random variables $x_i,x_j$ are asymptotically independent for $i\neq j$. 
 
For $a>a_c^{(I)}$, there is no solution to \eqref{eq:syst_a_resoudre0a}-\eqref{eq:syst_a_resoudre0b}, and the grand canonical approach fails. This typically signals a condensation transition. 
As we shall see in section~\ref{sec:condensed}, and similarly to what happens without disorder, one site takes an excitation of size $O(\sqrt{n})$, while the others remain of order $1$. Without disorder, the transition is for $a_c=2$. Hence the disorder modifies the transition, and, for some
distribution of the $\alpha_i$, may suppress it: if $\mathbb{E}_\alpha\left[ \frac{1}{\alpha^2}\right]=+\infty$, the condensed phase disappears.

\subsection{Models II and III}
The computations for models II and III are similar: the transition line is obtained by solving the grand canonical ensemble for $\mu=0$. 

For model II, Eqs. \eqref{eq:syst_a_resoudre1}-\eqref{eq:syst_a_resoudre2} become for $\mu=0$ and the parameter $a$ taking its critical value $a_c$:
\begin{eqnarray}
\frac1n \sum_{i=1}^n \frac{\int_0^\infty xe^{-\lambda x}dx}{z(\lambda,0)} &=& 1 
\label{eq:syst_a_resoudre1b} \\
\frac1n \sum_{i=1}^n \beta_i \frac{\int_0^\infty x^2e^{-\lambda x}dx}{z(\lambda,0)} &=& a_c.
\label{eq:syst_a_resoudre2b}
\end{eqnarray}
Eq.\eqref{eq:syst_a_resoudre1b} imposes $\lambda=1$; then, provided that $\mathbb{E}_\beta[\beta]$ exists, Eq.\eqref{eq:syst_a_resoudre2b} reads in the infinite $n$ limit $a_c=2\mathbb{E}_\beta[\beta]$. In the case without disorder $\beta={\rm cst}$, the transition point is $a_c=2\beta$. Hence for model II the transition point is not modified by the disorder, as soon as the expectation of $\beta$ is finite.
 
For model III, Eqs. \eqref{eq:syst_a_resoudre1}-\eqref{eq:syst_a_resoudre2} become for $\mu=0$ and the parameter $a$ taking its critical value $a_c$:
\begin{eqnarray}
\frac1n \sum_{i=1}^n \frac{\int_0^\infty xe^{-\lambda x}dx}{z(\lambda,0)} &=& 1 
\label{eq:syst_a_resoudre1c}\\
\frac1n \sum_{i=1}^n \left(\omega_i \frac{\int_0^\infty xe^{-\lambda x}dx}{z(\lambda,0)} +\frac{\int_0^\infty x^2e^{-\lambda x}dx}{z(\lambda,0)}\right) &=& a_c.
\label{eq:syst_a_resoudre2c}
\end{eqnarray}
If $\mathbb{E}_\omega[\omega]$ is finite, one obtains in the infinite $n$ limit $a_c=2+\mathbb{E}_\omega[\omega]$. In this case also, the transition point is not modified by the disorder, provided that $\mathbb{E}_\omega[\omega]$ is finite.

\section{The condensed phase, condensate localization}
\label{sec:condensed}
We give now more details on the condensed phase, and address the question: does the condensate (or the breather in DNLS words) localize on a specific site, or several specific sites?

The structure of the condensed phase in homogeneous models has been rigorously established in several cases, see for instance \cite{Grosskinsky03,Ferrari07} in a setting with one constraint, or \cite{Chaterjee10} for two constraints, as in this article. In the setting with two constraints the detailed studies \cite{Szavits14a,Szavits14b} rely on large deviations results for identically distributed random variables \cite{Nagaev69}. There are also a number of rigorous studies on heterogeneous, or disordered, models; these studies usually aim at describing stationary measures of particles systems with one conservation law (the number of particles), hence they fit in the "one constraint" setting \cite{Ferrari07,Grosskinsky11,Chleboun14,Mailler16}. None of these results apply directly to our case, and we are not aware of any rigorous study on the disordered, two constraints, setting.
Nevertheless, there is a natural assumption for the condensed phase: when $m_2/m_1^2>a_c$,  
the overwhelmingly most probable configuration corresponds to all $x_i$s except one being distributed according to the grand canonical distribution with parameters 
$(\lambda=\lambda^\ast(m_1,m_1^2a_c),\mu=0)$: hence, they are asymptotically independent, and the marginal distribution of $x_i$ is an exponential law with parameter 
$\lambda^\ast(1,a_c) \alpha_i$ (model I) or $\lambda^\ast(1,a_c)$ (models II and III). The last random variable $x_{i_0}$ absorbs the "excess second moment", taking the large value $m_1\sqrt{n}\sqrt{a-a_c}(1+o(1))$ (model I and III) or $(m_1/\beta_{i_0})\sqrt{n}\sqrt{a-a_c}(1+o(1))$ (model II). 
We will consider this picture as a reasonable assumption, which will be numerically confirmed, but waiting for a more rigorous justification. We now want to understand how is selected the variable which takes the large $O(\sqrt{n})$ value.

\subsection{Model I}
\label{sec:modelI}
We start again with model I, and assume $m_1=1$, $m_2>a_c=2\mathbb{E}_\alpha\left[1/\alpha^2\right]$. We call $q^I_i$ the probability that the condensate sits 
on site $i$.
First, we need more information on the size of the condensate, and how it depends on the site $i$ on which it resides. 
Assuming the condensate is on site $i$, all $N-1$ variables $x_j,~j\neq i$ are distributed according to an exponential law $\mathcal{E}(\alpha_j)$.
Hence $\mathbb{E}_\alpha(x_j^2)=2\alpha_j^{-2}$, which is finite by hypothesis. 
Hence the law of large numbers ensures for large $n$
\begin{equation}
\sum_{j\neq i} x_j^2 = 2n\mathbb{E}_{\alpha}(\alpha^{-2}) +o(n) = na_c +o(n). 
\label{eq:sumx2}
\end{equation}
We conclude that the condensate's size does not depend on $i$ at leading order, and is: 
\begin{equation}
x_i=\sqrt{n}\sqrt{a-a_c} + o(\sqrt{n})
\label{eq:size}
\end{equation}
A comparison with numerical simulations, using the algorithm described in appendix~\ref{sec:simulations}, is shown on Fig.\ref{fig:size}.
We add two remarks:\\
i) If, for some $\delta>0$, $\mathbb{E}_\alpha(\alpha^{-4-\delta})<+\infty$, the central limit theorem applies, with the usual $\sqrt{n}$ scaling, to the sum in \eqref{eq:sumx2} (this comes from the Lyapunov criterion, see for instance \cite{Billingsley}, theorem 27.3): the $o(n)$ in \eqref{eq:sumx2} term then becomes a $O(\sqrt{n})$, with gaussian distribution, and the $o(\sqrt{n})$ in \eqref{eq:size} becomes a $O(1)$, still with gaussian distribution (all this is at leading order in $n$); this is not the case if $\mathbb{E}_\alpha(\alpha^{-4})=+\infty$. Hence, depending on the distribution of the disorder, we can distinguish two regimes for the fluctuations in the condensate's size \footnote{We leave open here the limit case where $\forall \delta>0$ $\mathbb{E}_\alpha(\alpha^{-4-\delta})=+\infty$ and $\mathbb{E}_\alpha(\alpha^{-4})<+\infty$; we will also exclude the similar limit cases for models II and III.}. In the following, we assume for simplicity $\mathbb{E}_\alpha(\alpha^{-4-\delta})<+\infty$ (normal fluctuations).
\\
ii) Beyond the fluctuating term in \eqref{eq:size}, there may be bias term, which a priori depends on $i$; it will not enter at a relevant order in the following.\\  

\begin{figure}
\centerline{\includegraphics[width=12cm]{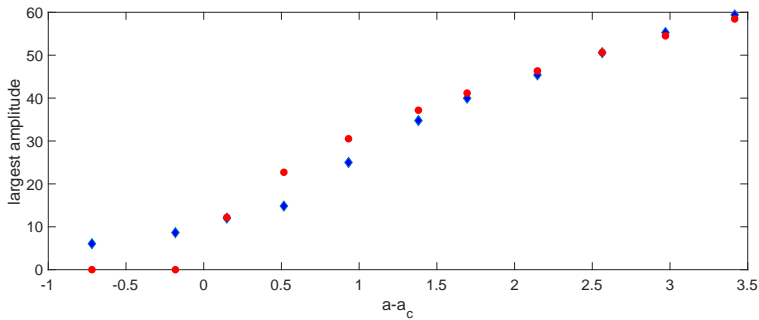}}
\caption{Largest $x_i$ variable vs $a-a_c$, for model I. Blue diamonds are numerical values, obtained with $n=1000$; the largest $x_i$ is recorded 
and averaged over $4.10^4$ MC sweeps. Each point corresponds to a single realization of the disorder. See appendix~\ref{sec:simulations} for a description of the algorithm.
Red circles are from formula \eqref{eq:size}, which is supposed to be valid only for $a-a_c>0$, and only at leading order. Strong finite size effects make it difficult to read the transition point from the numerics. For $a-a_c$ large enough, formula \eqref{eq:size} is satisfactory.
\label{fig:size}}
\end{figure}

\begin{figure}
\centerline{\includegraphics[width=12cm]{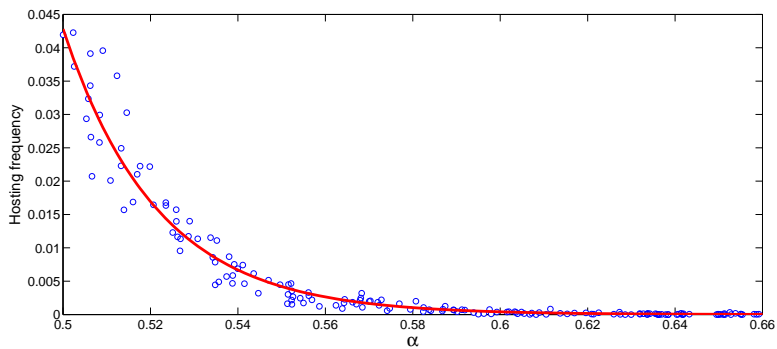}}
\caption{Histogram of the breather position as a function of $\alpha$. Simulation with $n=1000$ sites, and the law of $\alpha$ is uniform over $[0.5,1.5]$. The parameter $a=5>a_c$, well into the breather region. The breather position is recorded each $10$ MC sweeps to build the histogram. There is no averaging over the disorder: a single disorder realization is used. The red curve is the theoretical prediction. Only the smallest $\alpha$ are represented. \label{fig1}}
\end{figure}

We can now write that the probability that a condensate with this size indeed sits on site $i$ is proportional to the volume accessible to the other sites, with the constraint induced by 
\eqref{eq:size} (since we assume normal fluctuations, the $o(\sqrt{n})$ is actually a $O(1)$). This yields
\begin{equation}
\begin{split}
q^I_i \propto &{\rm Vol}\left( \left\{ (x_j\geq 0)_{j\neq i}~,~\sum_{j\neq i} \alpha_j x_j = n-\sqrt{n}\alpha_i\sqrt{a-a_c}+O(1)~,\right. \right.\\ 
&\left. \left.~\sum_{i\neq j} x_j^2 = a_cn +O(\sqrt{n})\right\} \right),
\end{split}
\label{eq:qi}
\end{equation}
where ${\rm Vol}(\cdot)$ stands for the volume.
Let us now define
\begin{eqnarray}
\Sigma_1^{\mathbf{\alpha}}(m_1,N) &=&  \left\{ (x_k)_{k= 1}^N~,~x_k\geq 0~,~\sum_{k} \alpha_k x_k \in [Nm_1-C_1,Nm_1+C_1] \right\} \nonumber \\
\Sigma_2(m_2,N) &=&  \left\{ (x_k)_{k= 1}^N~,~x_k\geq 0~,~\sum_{k}  x_k^2 \in [Nm_2-C_2\sqrt{N},Nm_2+C_2\sqrt{N}] \right\}, \nonumber
\end{eqnarray}
where $C_1$ and $C_2$ are constants. Clearly, the volume of $\Sigma_i$ depends on $C_i$, but this will be of no consequence. 
In order to compute $q_i$, we would like to estimate ${\rm Vol}\left(\Sigma_1^{\mathbf{\alpha}}(m_1,N) \cap \Sigma_2(m_2,N)\right)$, for appropriate $m_1$, $m_2$ and $N$. A simple computation yields (see appendix 2):
\begin{equation}
{\rm Vol}\left( \Sigma_1^{\mathbf{\alpha}}(m_1,N) \right) = \tilde{C}_1(1+o(1)) \frac{(Nm_1)^{N}}{\left(\Pi_i\alpha_i\right)N!},
\label{eq:vol1}
\end{equation}
with $\tilde{C}_1$ a constant depending only on $C_1$, and not on $N$.
Now, we notice that if the $x_k$ are picked up uniformly at random in $\Sigma_1^{\mathbf{\alpha}}(m_1,N)$, then they are asymptotically independent and distributed according to exponential laws $\mathcal{E}(\alpha_k/m_1)$ when $N$ tends to infinity. 
Thus, $(\sum_k x_k^2)/N$ tends to $m_1^2a_c$, by the law of large numbers, with fluctuations of order $1/\sqrt{N}$ (recall that $a_c =2\mathbb{E}_\alpha(\alpha^{-2})$, and we assume $\mathbb{E}_\alpha(\alpha^{-4-\delta})<+\infty$). Hence we conclude that a random point in 
$\Sigma_1^{\mathbf{\alpha}}(m_1,N)$ has a finite probability to be also in $\Sigma_2(m_1^2a_c,N)$, and  
\[
{\rm Vol}\left(\Sigma_1^{\mathbf{\alpha}}(m_1,N) \cap \Sigma_2(m_1^2a_c,N) \right) =O(1){\rm Vol} \left( \Sigma_1^{\mathbf{\alpha}}(m_1,N)\right)
\]
Using this result and \eqref{eq:vol1} for $N=n-1$, $m_1=1-n^{-1/2}\alpha_i \sqrt{a-a_c} +O(n^{-1})$ and $m_2=a_c+O(n^{-1/2})$, we obtain
\begin{eqnarray}
q^I_i &\propto & O(1) \frac{(n-1)^{n-1}\left(1-\frac{1}{\sqrt{n}}\alpha_i \sqrt{a-a_c}+O(\frac1n)\right)^{n-1}}{(n-1)!} \nonumber \\
\label{eq:qifinal}
\end{eqnarray}
where we have used Stirling formula. The $O(1)$ factor may a priori depend on $i$, but
since all $\alpha_i$ are strictly positive, and $a>a_c$, the $e^{-\sqrt{n(a-a_c)}\alpha_i}$ is the most important factor in \eqref{eq:qifinal} for large $n$. This shows the following:
\begin{enumerate}
\item The condensate has a tendency to localize on the sites with the smallest $\alpha$s.
\item However, in general the condensate does not select a single site (which would be the one with the smallest $\alpha$) in the large $n$ limit ; rather all sites with $\alpha_i$ within $O(1/\sqrt{n})$ of the smallest one have a sizable probability to host the condensate. 
\end{enumerate}
The number of sites with a sizable probability to host the condensate depends on the distribution of the $\alpha_i$. For a uniform distribution over an interval
$[\alpha_m,\alpha_M]$, there are typically $O(\sqrt{n})$ of them able to host the condensate. However, taking a distribution with less weight close to its minimum, it is possible to pin the condensate on a single site.
These conclusions are illustrated on Fig.\ref{fig1}. 

\subsection{Model II}
\label{sec:modelII}
\begin{figure}
\centerline{\includegraphics[width=12cm]{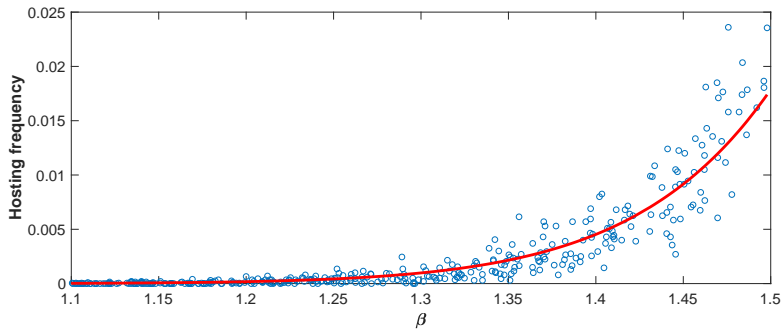}}
\caption{Histogram of the breather position as a function of $\beta$. Simulation with $n=1000$ sites, and the $\beta$ uniformly distributed over $[0.5,1.5]$. The parameter $a=4.29>a_c$, well into the breather region. The breather position is recorded each $10$ MC sweeps to build the histogram. There is no averaging over the disorder: a single disorder realization is used. The red curve is the theoretical prediction \eqref{eq:probaII}. Only the sites with $\beta>1.1$ are shown. \label{fig2}}
\end{figure}

We assume here that the disorder distribution is such that $\forall i$, $\beta_i>0$ and $\mathbb{E}_\beta(\beta_i)=1$. We also assume $m_1=1$, and the system is in the condensed phase: $m_2 > a_c=2$. The picture is then similar as the one for model I: $n-1$ variables are asymptotically independent and distributed according to
the same exponential law $\mathcal{E}(1)$; the last variable, say $x_i$, hosts the condensate and is equal at leading order to $\sqrt{n(m_2-a_c)/\beta_i}$.
Our goal is to determine $q_i^{II}$, the probability that the condensate is hosted on site $i$.
As for model I, we need to know the size of the condensate more precisely. We have
\begin{equation}
\label{eq:sum22}
\sum_{j\neq i} \beta_j x_j^2 =2 \sum_{j\neq i}\beta_j + O(\sqrt{n}) +O_i(1)= 2 n\mathbb{E}_\beta(\beta)  + o(n);
\end{equation}
as above, we can distinguish between a normal fluctuation regime, when $\mathbb{E}(\beta^{2+\delta})<+\infty$ for some $\delta>0$, in which case the $o(n)$ term above becomes a $O(\sqrt{n})$, and an anomalous fluctuation regime, when $\mathbb{E}(\beta^2)=+\infty$. We assume for simplicity in the following that
$\mathbb{E}(\beta^{2+\delta})<+\infty$ for some $\delta>0$. The leading order in the remainder $O(\sqrt{n})$ term does not depend on $i$, but higher orders do.
From \eqref{eq:sum22}, we obtain the size of the condensate:
\begin{equation}
\label{eq:size2}
\beta_i x_i^2 = n(m_2-2) + O(\sqrt{n}) ~,~x_i = \sqrt{\frac{n(m_2-2)}{\beta_i}} + O_i(1).
\end{equation}
Notice that in this case the size of the condensate depends on its location, and so does the leading order fluctuating correction in the second equation of \eqref{eq:size2}, which is hence denoted $O_i(1)$. We define
\begin{eqnarray}
\Sigma_1(m_1,N) &=&  \left\{ (x_k)_{k= 1}^N~,~x_k\geq 0~,~\sum_{k} x_k \in [Nm_1-C_1,Nm_1+C_1] \right\} \nonumber \\
\Sigma_2^{{\mathbf \beta}}(m_2,N) &=&  \left\{ (x_k)_{k= 1}^N~,~x_k\geq 0~,~\sum_{k}  \beta_k x_k^2 \in [Nm_2-C_2\sqrt{N},Nm_2+C_2\sqrt{N}] \right\}. \nonumber
\end{eqnarray}
$q_i^{II}$ is then proportional to the phase space volume available to the other variables $(x_j)_{j\neq i}$, when $x_i$ is fixed, up to fluctuations, to the condensate value:
\[
q_i^{II} \propto {\rm Vol} \left[\Sigma_1(1-\sqrt{(m_2-2)/(n\beta_i)},n-1) \cap \Sigma_2^{{\mathbf \beta}}(2,n-1)\right].
\]  
The reasoning is as in \ref{sec:modelI}: for a point in $\Sigma_1$, the constraint represented by $\Sigma_2$ is typically satisfied. Hence it is enough to compute the 
volume of $\Sigma_1$, which is done using the appendix. We obtain
\begin{equation}
q_i^{II} \propto K(n) e^{-\sqrt{\frac{n(m_2-2)}{\beta_i}}},
\label{eq:probaII}
\end{equation}
where the prefactor $K(n)$ a priori depends on $i$, as a consequence of the $O_i(1)$ correction in \eqref{eq:size2}; the dominant term is still given by the exponential. We conclude:
\begin{enumerate}
\item The condensate has a tendency to localize on the sites with the largest non linearity $\beta$.
\item However, as in \ref{sec:modelI}, in general the condensate does not select a single site (which would be the one with the largest $\beta$) in the large $n$ limit; rather all sites with $\beta_i$ within $O(1/\sqrt{n})$ of the largest one have a sizable probability to host the condensate. 
\end{enumerate}
These conclusions are illustrated on Figure~\ref{fig2}.

\subsection{Model III}
\begin{figure}
\centerline{\includegraphics[width=12cm]{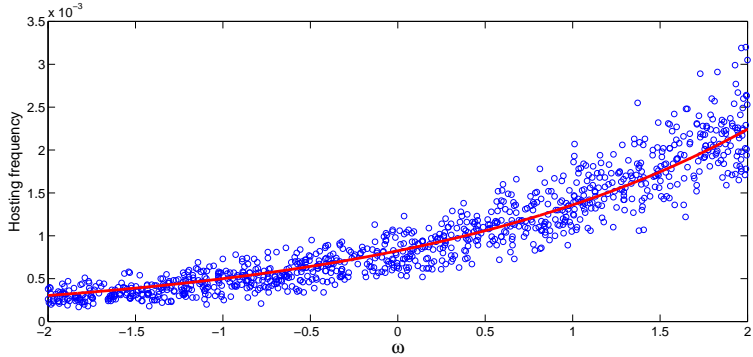}}
\caption{Histogram of the breather position as a function of $\omega$. Simulation with $n=1000$ sites, and the $\omega$ uniformly distributed over $[-2,2]$. The parameter $a=4.65>a_c$, well into the breather region. The breather position is recorded each $10$ MC sweeps to build the histogram. There is no averaging over the disorder: a single disorder realization is used. The red curve is the theoretical prediction \eqref{eq:probaIII}. Contrary to Figs.~\ref{fig1} and \ref{fig2}, all values of $\omega$ are represented.\label{fig3}}
\end{figure}

We assume here that the disorder distribution is such that the $\omega_i$ are identically distributed, with zero expectation. We also assume $m_1=1$, and the system is in the condensed phase: $m_2 > a_c=2$. The picture is then similar to the one for models I and II: $n-1$ variables are asymptotically independent and distributed according to the same exponential law $\mathcal{E}(1)$; the last variable, say $x_i$, hosts the condensate and is equal at leading order to $\sqrt{n(m_2-2)}$. Our goal is now to determine $q_i^{III}$, the probability that the condensate is hosted on site $i$.
It is now necessary to compute the size of the condensate beyond leading order. Let us first assume that
$\mathbb{E}_\omega(\omega^{2+\delta})<+\infty$ for some $\delta>0$.  In this case the central limit theorem applies to the first sum in the following equation: 
\begin{equation}
\sum_{j\neq i}\omega_j x_j + \sum_{j\neq i} x_j^2 =  \sqrt{2} \sigma_\omega \sqrt{n} Z + 2n + \sqrt{2} \sqrt{n} \tilde{Z}+ O_i(1),
\label{eq:sumxIII}
\end{equation}
where $Z,\tilde{Z}$ are normalized gaussian variables, $\mathbb{E}_\omega(\omega^2)=\sigma_\omega^2<+\infty$, and the $O_i(1)$ term depends on $i$.
We obtain the following equation for the condensate $x_i$:
\[
\omega_i x_i +x_i^2 = nm_2-2n+ O(\sqrt{n}) +O_i(1).
\]
One finds at leading order $x_i=\sqrt{n(m_2-2)}$ as anticipated, and we need to go further to understand the dependency on $i$:
\begin{equation}
x_i = \sqrt{n(m_2-2)} -\frac12 \omega_i +O(1) +O_i(\frac{1}{\sqrt{n}}).
\end{equation}
Notice that the second and third term are of the same order of magnitude; however, we know that the $O(1)$ is a fluctuating term which does not depend on $i$. We define
\begin{eqnarray}
\Sigma_3^{{\mathbf \omega}}(m_2,N) &=&  \left\{ (x_k)_{k= 1}^N~,~x_k\geq 0~,~\sum_{k}  (\omega_k x_k +x_k^2) \in [Nm_2-C_2\sqrt{N},Nm_2+C_2\sqrt{N}] \right\}. \nonumber
\end{eqnarray}
Then
\[
q_i^{III} \propto {\rm Vol} \left[\Sigma_1(1-\sqrt{(m_2-2)/n}+\frac12 \frac{\omega_i}{n} +O(\frac1n),n-1) \cap \Sigma_3^{{\mathbf \omega}}(2,n-1)\right].
\]  
Again, for a point in $\Sigma_1$, the constraint represented by $\Sigma_3^{{\mathbf \omega}}$ is typically satisfied. Computing the volume of $\Sigma_1$, we obtain
\begin{equation}
q_i^{III} \propto K(n) e^{\frac{\omega_i}{2}},
\label{eq:probaIII}
\end{equation}
where the prefactor $K(n)$ does not depend on $i$ at leading order.
We conclude:
\begin{enumerate}
\item The condensate has a tendency to localize on the sites with the largest on site frequencies $\omega$s.
\item However, this tendency is rather weak, as it does not depend on $n$: typically, all sites have a sizable probability to host the condensate. 
\end{enumerate}
These conclusions are illustrated on Fig.\ref{fig3}.
Finally, if $\mathbb{E}_\omega(\omega^2)=+\infty$, the central limit theorem scaling $\sqrt{n}$ for the first term in the rhs of \eqref{eq:sumxIII} is not valid anymore; the size of the condensate is still at leading order $\sqrt{n(m_2-2)}$, but the first correction is a fluctuating 
term which does not depend on $i$. 

\section{Conclusion}

We first recall our main findings: i) the phase diagram corresponding to the homogenous case \eqref{eq:1} easily generalizes to the disordered case when the disorder is weak enough: the transition point may be shifted; if the disorder is strong enough (ie with a wide enough distribution), the condensed phase disappears. 
ii) the condensate localization undergoes a "partial symmetry breaking": the choice of the site hosting the condensate is not determined by the disorder, but merely biased by it. To be more precise, and using the vocabulary of the disordered DNLS equation, the sites with highest onsite nonlinearity, or highest on site frequency, are more likely to host the condensate.

Clearly, several open problems remain. First, our description of the condensed phase relies on heuristic arguments and Monte Carlo simulations; a rigorous description is lacking. Second, models I, II and III provide an idealized picture of  the disordered DNLS system; what happens when coupling between sites is taken into account? With the homogeneous DNLS case in mind, we expect the main features seen here (the transition, and the partial symmetry breaking regarding the condensate localization) to hold at least qualitatively in presence of coupling. Simulations, or a better theory, are however needed. 
We end by mentioning \cite{Juanico07}:  this article studies the formation of localized excitations in a model of a protein, and remarks that these breather-like localized modes are more likely to form at the stiffest parts of the protein, which correspond in our language to largest on-site frequencies. This is an encouraging sign towards the applicability of the concepts of this article to more realistic models, but there is obviously a lot of work to do to prove, or disprove, the connection with \cite{Juanico07}. 

\section{Appendix 1}
\label{sec:appendix1}
We have to show that the maximum of $\varphi_2$ on the curve $\varphi_1=1$ is attained at $\mu=0$. Let us call $\tilde{\varphi}_2(\mu)$ the function
$\varphi_2$ along the curve $\varphi_1(\lambda,\mu)=1$. It is enough to show that $d\tilde{\varphi}_2/d\mu \leq 0$. From the implicit function theorem, 
we have
\[
\frac{d\tilde{\varphi}_2}{d\mu} =  \frac{-\partial_\lambda \varphi_2\partial_\mu \varphi_1 +\partial_\lambda \varphi_1 \partial_\mu \varphi_2}{\partial_\lambda \varphi_1}.
\]
We introduce the notation $<\cdot>_{\alpha}$:
\[
<f(x)>_{\alpha} =\frac{\int_0^{+\infty} f(x) e^{-\lambda \alpha x-\mu x^2} dx}{z(\lambda\alpha,\mu)}
\]
Note that the result of this average depends on $\alpha$. 
Then
\begin{eqnarray}
\partial_\lambda \varphi_1 &=& \frac1n \sum_{i=1}^n \left(<\alpha_i x>_{\alpha_i}^2-<\alpha_i^2 x^2>_{\alpha_i}\right) \underset{n\to \infty}{\longrightarrow} 
\mathbb{E}_\alpha \left[<\alpha  x>_{\alpha}^2-<\alpha^2 x^2>_{\alpha}\right]\nonumber \\
\partial_\mu \varphi_1 &=& \frac1n \sum_{i=1}^n \left(<\alpha_i x>_{\alpha_i}<x^2>_{\alpha_i}-<\alpha_i x^3>_{\alpha_i}\right) \underset{n\to \infty}{\longrightarrow} 
\mathbb{E}_\alpha \left[<\alpha x>_{\alpha}<x^2>_{\alpha}-<\alpha x^3>_{\alpha}\right]\nonumber \\
\partial_\lambda \varphi_2 &=& \frac1n \sum_{i=1}^n \left(<\alpha_i x>_{\alpha_i}<x^2>_{\alpha_i}-<\alpha_i x^3>_{\alpha_i}\right)\underset{n\to \infty}{\longrightarrow} 
\mathbb{E}_\alpha \left[<\alpha x>_{\alpha}<x^2>_{\alpha}-<\alpha x^3>_{\alpha}\right] \nonumber \\
\partial_\mu \varphi_2 &=& \frac1n \sum_{i=1}^n \left(<x^2>_{\alpha_i}^2-<x^4>_{\alpha_i}\right) \underset{n\to \infty}{\longrightarrow} 
\mathbb{E}_\alpha \left[<x^2>_{\alpha}^2-<x^4>_{\alpha}\right],\nonumber 
\end{eqnarray}
where $\mathbb{E}_\alpha$ stands for the expectation with respect to the quenched disorder, and the limits are consequences of the law of large numbers.
Finally, we obtain 
\begin{eqnarray}
\frac{d\tilde{\varphi}_2}{d\mu} &\underset{n\to \infty}{\longrightarrow}& \frac{\left(\mathbb{E}_\alpha\left[ \langle (\alpha x-<\alpha x>_\alpha)(x^2-<x^2>_\alpha)\rangle_\alpha\right]\right)^2 - \mathbb{E}_\alpha\left[\langle(\alpha x-<\alpha x>_\alpha)^2\rangle_\alpha\right] \mathbb{E}_\alpha\left[\langle (x^2-<x^2>_\alpha)^2\rangle_\alpha\right]}{\mathbb{E}_\alpha \left[\langle(\alpha x-<\alpha x>_{\alpha})^2\rangle_\alpha\right]} \nonumber \\
&\leq& 0, \nonumber 
\end{eqnarray}
where the last line is from Cauchy-Schwarz inequality. Hence the maximum of $\varphi_2$ on the curve $\varphi_1=1$ is attained at $\mu=0$. From the constraint 
$\varphi_1=1$ and $\mu=0$, it is easy to see that $\lambda=1$; then the sought maximum can be computed:
\[
\underset{\mu,\varphi_1(\lambda,\mu)=1}{\rm max} \varphi_2(\lambda,\mu) = \frac1n \sum_{i=1}^n \frac{1}{\alpha_i^2} \underset{n\to \infty}{\longrightarrow} 2\mathbb{E}_\alpha\left[\frac{1}{\alpha^2}\right],
\]
where the last line requires that $\mathbb{E}_\alpha(1/\alpha^2)$ is finite.

\section{Appendix 2}
\label{sec:appendix2}
We start with the simple remark:
\[
{\rm Vol}\left( \left\{ (x_k)_{k= 1}^N~,~x_k\geq 0~,~\sum_{k} x_k\leq 1 \right\}\right) =\frac{1}{N!}.
\]
From this we can easily compute the volume of the set:
\[
\Sigma_1(m_1,N) =  \left\{ (x_k)_{k= 1}^N~,~x_k\geq 0~,~\sum_{k}  x_k \in [Nm_1-C_1,Nm_1+C_1] \right\},
\]
where $m_1,C_1$ are order $1$ and $N$ is large: this set is a slightly thickened surface. 
We obtain
\[
{\rm Vol}\left( \Sigma_1(m_1,N) \right) = \frac{1}{N!} [(Nm_1+C_1)^N-(Nm_1-C_1)^N] =\frac{(Nm_1)^N}{N!} \left(2\sinh \frac{C_1}{m_1}+O\left(\frac1N\right)\right). 
\]
We note that $\Sigma_1$ depends on $C_1$ only through the prefactor, which will not be important for our computations.
A further change of variables provides
\[
{\rm Vol}\left( \Sigma_1^\alpha(m_1,N) \right) = \frac{1}{N!} [(Nm_1+C_1)^N-(Nm_1-C_1)^N] =\frac{(Nm_1)^N}{N!\Pi_k \alpha_k} \left(2\sinh \frac{C_1}{m_1}+O\left(\frac1N\right)\right).
\]

\section{Appendix 3}
\label{sec:simulations}

We have to sample points uniformly on the set (model I):
\begin{eqnarray}
\begin{array}{ll}
{\rm Model~I} & \begin{cases}\forall i=1,\ldots,n~,~x_i \geq 0 \\ \sum_{i=1}^n \alpha_i x_i = n m_1\\ \sum_{i=1}^n x_i^2 = n m_2  \end{cases}
\end{array}
\label{eq:model1b}
\end{eqnarray}
We use for this purpose a Monte Carlo algorithm, relying on the following Markov Chain. At each step, 3 different indices are drawn uniformly from $\{1,\ldots,n\}$, and denoted $i_1,i_2,i_3$; we write $\alpha_{i_1}x_{i_1}+\alpha_{i_2}x_{i_2}+\alpha_{i_3}x_{i_3} = s_1$, and $x_{i_1}^2+x_{i_2}^2+x_{i_3}^2=s_2$. The set
\[
\{ \alpha_{i_1}y_1+\alpha_{i_2}y_2+\alpha_{i_3}y_3 = s_1~,~y_1^2+y_2^2+y_3^2 =s_2 \}
\]
is a circle, containing $(x_{i_1}, x_{i_2},x_{i_3})$. By construction replacing $(x_{i_1}, x_{i_2},x_{i_3})$ by any point $(y_1, y_2,y_3)$ on the circle provides a new configuration $(x_1,\ldots,x_n)$ which satisfies the two equality constraints in \eqref{eq:model1b}. We now pick up uniformly at random a point $(y_1, y_2,y_3)$
on this circle, and accept the move $(x_{i_1}, x_{i_2},x_{i_3}) \to (y_1, y_2,y_3)$ if $y_1,y_2,y_3\geq 0$. In our simulations the rate of acceptance is typically about $50$ percent. The equilibrium measure of this Markov chain is what we are looking for. Hence, running it for a sufficiently long time yields a point approximately uniformly sampled from \eqref{eq:model1b}. This simulation strategy is rather natural; it is a straightforward generalization of the algorithm used for instance in \cite{Iubini13,Iubini14,Iubini17,Szavits14b} in cases without disorder.

The above algorithm can be used almost without changes for model III. The main part is to sample uniformly on the set
\[
\{ y_1+y_2+y_3 = s_1~,~\omega_1y_1+\omega_2y_2+\omega_3y_3 + y_1^2+y_2^2+y_3^2 =s_2 \},
\] 
which is still a circle. For model II, we have to sample uniformly on the set
\[
\{ y_1+y_2+y_3 = s_1~,~\beta_1 y_1^2+\beta_2y_2^2+\beta_3y_3^2 =s_2 \},
\] 
which is an ellipse. This is a bit more technical, but does not pose any major difficulty.

Our simulations typically use $n=1000$ variables. They start with a relaxation run of $10^3$ Monte Carlo sweeps (that is $10^6$ MC steps), before we start recording points.\\


\end{document}